\documentclass[aps,prc,preprint,showpacs]{revtex4}
\usepackage{epsfig}
\usepackage{amsmath}
\usepackage[hyperref]{hyperref}

\begin{document}

\title{Neutron contribution to nuclear DVCS asymmetries}

\author{V. Guzey}
\email{vguzey@jlab.org}
\homepage{http://www.jlab.org/~vguzey}
\affiliation{Theory Center, Jefferson Lab, Newport News, VA 23606, USA}

\begin{abstract}

Using a simple model for nuclear GPDs, we study
the role of the neutron contribution to nuclear DVCS observables. As an example,
we use the beam-spin asymmetry $A_{LU}^A$ measured in coherent and 
incoherent DVCS on a wide range of nuclear targets in the 
HERMES and JLab kinematics. 
We find that at small values of the momentum transfer $t$, 
$A_{LU}^A$ is dominated by the  coherent-enriched contribution, 
which enhances $A_{LU}^A$ compared to the free proton asymmetry
$A_{LU}^p$, $A_{LU}^A(\phi)/A_{LU}^p(\phi)=1.8-2.2$.
At large values
of $t$, the nuclear asymmetry is dominated by the incoherent
contribution and $A_{LU}^A/(\phi)A_{LU}^p(\phi)=0.66-0.74$.
The deviation of $A_{LU}^A(\phi)/A_{LU}^p(\phi)$ from unity at large $t$ is a result of the
neutron contribution, which gives a possibility to constain neutron GPDs in incoherent
nuclear DVCS. A similar trend is expected for other DVCS asymmetries.

\end{abstract}


\preprint{JLAB-THY-08-767}

\maketitle
\section{Introduction}
\label{sec:Intro}

Hard exclusive reactions such as  Deeply Virtual Compton Scattering (DVCS),
$\gamma^{\ast}T \to \gamma T^{\prime}$, and hard exclusive meson production (HEMP),
$\gamma^{\ast}T \to M T^{\prime}$, have emerged as indispensable tools to 
access the microscopic (parton) properties of hadrons~\cite{Mueller:1998fv,Ji:1996nm,Ji:1998pc,Radyushkin:1996nd,Radyushkin:1997ki,Radyushkin:2000uy,Collins:1998be,Collins:1996fb,Brodsky:1994kf,Goeke:2001tz,Diehl:2000xz,Belitsky:2001ns,Diehl:2003ny,Belitsky:2005qn}.
In the above reactions, $T$ and $T^{\prime}$ stand for any hadronic target (nucleon, pion, atomic 
nucleus); $M$ denotes any meson. Note that the above reactions may also include
transitions between different hadronic states such as 
e.g.~$N \to \Delta$, $p \to n$, $N \to N \pi$~\cite{Mankiewicz:1998kg,Frankfurt:1999fp,Polyakov:2006dd}
  and production of pairs of mesons~\cite{Polyakov:1998ze}.  

In the Bjorken limit (large $Q^2$), 
the QCD factorization theorem for DVCS and HEMP on any hadronic 
target~\cite{Collins:1998be,Collins:1996fb} states that
corresponding scattering amplitudes factorize in
 convolution of perturbative (hard) coefficient
functions with nonperturbative (soft) matrix elements, 
which are parameterized in terms of generalized parton distributions 
(GPDs). GPDs are universal (process-independent) functions that contain information
on parton distributions and correlations in hadrons and in matrix elements
describing transitions between different 
hadrons (see above).

In this paper, we consider DVCS on nuclear targets, 
$\gamma^{\ast}A \to \gamma A$, which gives an access to nuclear GPDs.
We would like to single out the 
following three important roles of nuclear DVCS:
\begin{itemize}
\item It gives information on nucleon GPDs, which is complimentary
to that obtained in DVCS on the free proton;
\item It allows to study novel nuclear effects, which decouple from
DIS and elastic scattering on nuclei;
\item
It imposes stringent constraints on theoretical models attempting to give
a covariant description of the nuclear structure. 
\end{itemize}

In this paper, we deal with the first point. In particular, we 
examine the role of the neutron contribution to nuclear DVCS asymmetries on a wide
range of nuclei. This allows one to constrain neutron GPDs, which are
not directly accessible.

Nuclear DVCS opens possibilities to study novel nuclear effects,
which seem to be predominantly encoded in the real part of the DVCS scattering
amplitude. 
It was speculated in the framework of the nuclear liquid drop model that 
the so-called nuclear
$D$-term, which contributes to the real part of 
the nuclear DVCS amplitude, has a fast, non-trivial 
dependence on the atomic number $A$ 
($A^{7/3}$ vs.~naively expected $A^2$)~\cite{Polyakov:2002yz}.
This observation was confirmed by an
analysis of nuclear GPDs using the Walecka model~\cite{Guzey:2005ba}.
 In that analysis, 
the fast $A$-dependence of nuclear GPDs comes from nuclear meson degrees
of freedom. Hence, the measurement of DVCS observables sensitive to the
real part of the DVCS amplitude gives a possibility to study non-nucleon
(mesonic)
degrees of freedom in nuclei.

In the small Bjorken $x_B$ limit,
a model for nuclear GPDs, which combines the 
model for nucleon GPDs based on the aligned-jet model
with phenomenological parameterizations of usual nuclear PDFs,
was suggested in~\cite{Freund:2003wm,Freund:2003ix}.
It was found that
 the ratio of the real parts of the nuclear to nucleon DVCS
amplitudes has a very unexpected behavior as a function of $x_B$,
 which is very different from the corresponding ratio of the 
imaginary parts. The latter was found to be similar to the ratio of the 
nuclear to nucleon structure functions measured in inclusive DIS.
This, again, hints that novel nuclear effects might be lurking in the real
part of the nuclear DVCS amplitude.

The third role of nuclear DVCS is related to the fact that
nuclear GPDs, similarly to nucleon GPDs, should obey the fundamental properties
of polynomiality and positivity. In order to achieve these properties, 
theoretical models used to build nuclear GPDs must give a covariant
description of the nuclear structure, which imposes severe constraints on
the nuclear models. This problem was discussed in relation to 
modeling deuteron GPDs in~\cite{Cano:2003ju}.

The literature on nuclear DVCS and nuclear GPDs in not numerous
and can be readily comprehensively overviewed.

Originally, the formalism of deuteron GPDs was developed in~\cite{Berger:2001zb}.
The formalism of nuclear GPDs of any spin-0, spin-1/2 and spin-1 nuclei
was presented in~\cite{Kirchner:2003wt}. Assuming that nuclei are collections 
of free protons and neutrons, predictions for DVCS observables (asymmetries) were
made. In particular, in accord with the earlier result of~\cite{Guzey:2003jh}, 
it was predicted that the nuclear DVCS 
beam-spin asymmetry is enhanced compared to the free
proton asymmetry, $A_{LU}^A(\phi)/A_{LU}^p(\phi) \sim 5/3$, for spin-0 and spin-1/2 
nuclei.

Up to now, the main theoretical approach to dynamical models of nuclear GPDs 
is the convolution approximation, which assumes that nuclear GPDs
are given by convolution of unmodified or modified nucleon GPDs with the distribution
of nucleons in the nuclear target. The latter distribution is obtained from the
non-relativistic nuclear wave function.
Within the convolution approximation, there were
considered GPDs of such nuclei as 
deuterium~\cite{Cano:2002ph,Cano:2002tn,Cano:2003ju}, 
$^3$He~\cite{Scopetta:2004kj,Scopetta:2006wu}, 
$^4$He~\cite{Liuti:2005qj,Liuti:2005gi}, $^{20}$Ne and $^{76}$Kr~\cite{Guzey:2003jh},
a wide range of nuclei from $^{12}$C to $^{208}$Pb~\cite{Guzey:2005ba}
(in that analysis, besides nucleons, meson degrees of freedom were also
used in the convolution).

While the convolution approximation is reliable for $x_B > 0.1$, it is not applicable
for small $x_B$, where such coherent nuclear effects as nuclear shadowing and
antishadowing become important. A model of nuclear GPDs for heavy nuclei,
which takes into account nuclear shadowing and antishadowing, 
was proposed in~\cite{Freund:2003wm,Freund:2003ix}
(see also the discussion above).

Another important aspect of nuclear DVCS, at least from the practical point
of view, is the interplay between the coherent (the nucleus remains intact) and
incoherent (the nucleus excites or breaks up) contributions to
nuclear DVCS. This  was studied in~\cite{Guzey:2003jh} and
a general expression for nuclear DVCS asymmetries, which interpolates between the
coherent and incoherent regimes, was derived.
It was predicted that for the coherent contribution, in the kinematics of the 
HERMES experiment, the ratio of 
the nuclear ($^{20}$Ne and  $^{76}$Kr)
to the free proton beam-spin asymmetries is enhanced, 
$A_{LU}^A(\phi)/A_{LU}^p(\phi)  \approx 1.8$.
For the incoherent contribution, it was predicted that
 $A_{LU}^A(\phi)/A_{LU}^p(\phi)=1$, 
provided that the neutron contribution to the nuclear DVCS amplitude is
neglected. 

It is the main goal of the present work to go beyond this approximation and to
study the role of the neutron contribution in  coherent 
and incoherent nuclear DVCS observables (asymmetries).

On the experimental side,
initial measurements of nuclear DVCS were reported by the HERMES collaboration at DESY~\cite{Ellinghaus:2002zw}
and more data on nuclear DVCS at HERMES is expected~\cite{Ellinghaus:2007dw}.
The CLAS collaboration at Jefferson Lab recently reported a measurement of 
DVCS on deuterium with the aim to study
the neutron GPDs~\cite{Mazouz:2007vj}.  
It is planned that nuclear GPDs will be studied at Jefferson Lab at the present
6 GeV and the future 12 GeV energy of the electron beam.
At high energies, nuclear GPDs will be studied 
at the LHC in ultraperipheral nucleus-nucleus collisions, see e.g.~\cite{Frankfurt:2003qy},
and at the future Electron-Ion Collider.

This paper is organized as follows.
In Sect.~\ref{sec:NuclGPDs}, we explain our model of nuclear GPDs.
The interpolating formula between the coherent and incoherent regimes of
nuclear DVCS is derived in Sect.~\ref{sec:Interpolation}.
Predictions for the nuclear beam-spin DVCS asymmetry in HERMES and 
JLab kinematics, with an emphasis on the neutron contribution,
are presented in Sect.~\ref{sec:nucl_asymmetry}. We summarize and
discuss our results in Sect.~\ref{sec:SD}.

\section{Model for nuclear and nucleon GPDs}
\label{sec:NuclGPDs}

\begin{figure}[h]
\begin{center}
\epsfig{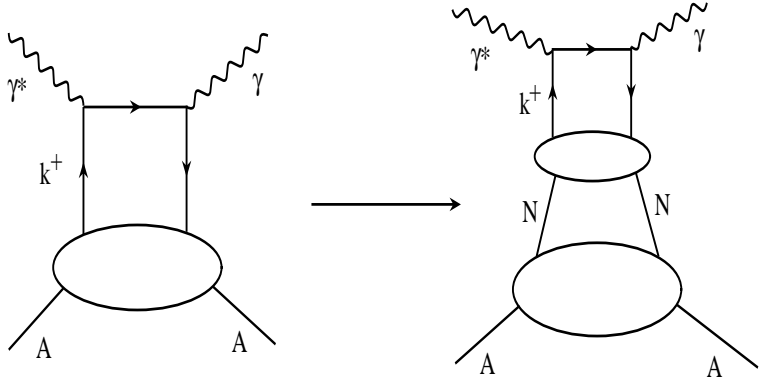}
\end{center}
\caption{Schematic representation of nuclear quark GPDs.}
\label{fig:model}
\end{figure}

We use a simple model for nuclear GPDs that captures main features of the dependence of 
nuclear GPDs on the atomic number $A$ and on the momentum transfer $t$.
We assume that the nucleus consists of $A$ uncorrelated nucleons: 
$Z$ protons and $N=A-Z$ neutrons~\cite{Kirchner:2003wt}, 
see Fig.~\ref{fig:model}.

For simplicity, we shall consider spin-0 nuclei. 
In this case, there is only one leading-twist quark nuclear GPD, $H_A^q$,
which can be expressed 
in terms of the free proton and neutron quark GPDs $H^q$ and $E^q$ as
follows,
\begin{eqnarray}
H^q_A(x,\xi_A,Q^2,t)&=&\left|\frac{dx_N}{dx}\right| \Bigg[Z\,\left(H^{q/p}(x_N,\xi_N,Q^2,t)+\frac{t}{4 m_N^2}E^{q/p}(x_N,\xi_N,Q^2,t)\right) \nonumber\\
&+&N\,\left(H^{q/n}(x_N,\xi_N,Q^2,t)+\frac{t}{4 m_N^2}E^{q/n}(x_N,\xi_N,Q^2,t)\right)
\Bigg] F_A(t) \,, 
\label{eq:model1}
\end{eqnarray}
where $F_A(t)$ is the nuclear form factor normalized to unity;
$m_N$ is the nucleon mass; other variables are introduced below.
Note that the GPDs $H^q$ and $E^q$ enter Eq.~(\ref{eq:model1}) in the 
combination that leads to the proper nuclear charge form factor~\cite{Guzey:2003jh}.

The Bjorken variable $x_A$ is defined with respect to the nuclear target.
In the laboratory frame, we have
\begin{equation}
x_A=\frac{Q^2}{2 \nu M_A}=\frac{Q^2}{2 \nu A m_N}=\frac{1}{A} \,x_B \,,
\label{eq:x}
\end{equation}
where $\nu$ is the photon energy; $M_A$ is the mass of the nucleus.
From the relations
\begin{equation}
\xi_A=\frac{x_A}{2-x_A} \,, \quad \xi_N=\frac{x_B}{2-x_B} \,,
\label{eq:xi}
\end{equation}
it follows that 
\begin{equation}
\frac{\xi_N}{1+\xi_N}=A\frac{\xi_A}{1+\xi_A} \,.
\label{eq:xiA}
\end{equation}

Next we find the relation between $x$ and $x_N$.
In the symmetric notation~\cite{Goeke:2001tz},
the outgoing interacting quark carries the plus-momentum $k^+=
(x+\xi_A) \bar{P}_A^+$, see the left-hand side of Fig.~\ref{fig:model}.
On the other hand, $k^+$ can also be written as  (see the right-hand side of Fig.~\ref{fig:model})
\begin{eqnarray}
k^+&=&(x_N+\xi_N)\bar{P}_N^{+}=(x_N+\xi_N)\left(\frac{1}{A}P_A^{+}+\frac{\Delta^+}{2}\right) 
\nonumber\\
&=&(x_N+\xi_N)\left(\frac{1}{A}(1+\xi_A)-\xi_A\right)\bar{P}_A^+ \,.
\label{eq:kplus}
\end{eqnarray}
In this derivation, we used the assumption that 
$P_N^{+}=P_A^{+}/A$.
Therefore, with help of Eq.~(\ref{eq:xiA}), we find that
\begin{equation}
\frac{x_N}{x}=\frac{\xi_N}{\xi_A} \,.
\label{eq:xA}
\end{equation}

In the forward limit, Eq.~(\ref{eq:model1}) reduces to the
model for 
 nuclear quark
parton distribution functions (PDFs),
\begin{equation}
q_A(x_A,Q^2)=A \left[Z\,q_p(x_B,Q^2)+N\,q_n(x_B,Q^2)\right]\,.
\label{eq:forward_limit}
\end{equation}
These nuclear PDFs satisfy the baryon number (total charge) and momentum 
sum rules, 
\begin{eqnarray}
&&\int^1_{-1} dx_A \sum_q e_q \,q_A(x_A,Q^2)=\int^1_{-1} dx_B \sum_q e_q  \left[Z\,q_p(x_B,Q^2)+N\,q_n(x_B,Q^2)\right]=Z \,, \nonumber\\
&&\int^1_{-1} dx_A \sum_q x_A q_A(x_A,Q^2)=\int^1_{-1} dx_B \sum_q x_B\left[\frac{Z}{A}\,q_p(x_B,Q^2)+\frac{N}{A}\,q_n(x_B,Q^2)\right] \,.
\end{eqnarray}

Taking the first $x$-moment of the nuclear GPD weighted with quark charges,
 one obtains the nuclear electric form factor,
\begin{equation}
F_A^{e.m.}(t)  \equiv 
\int^1_{-1} dx \sum_q e_q\,H^q_A(x,\xi_A,Q^2,t)
=\left[Z F_{E}^p(t)+NF_{E}^n(t)\right]F_A(t) \,,
\label{eq:Fem}
\end{equation}
where $F_E^{p,n}(t)=F_1^{p,n}(t)+t/(4 m_N^2) F_2^{p,n}(t)$ are the electric
form factors of the proton and neutron expressed in terms 
of the corresponding Dirac and Pauli form factors.

The fact that the right-hand side of Eq.~(\ref{eq:Fem}) does not depend
on $\xi_A$ means that the first $x$-moment of $H^q_A$ satisfies polynomiality.
An examination shows that higher $x$-moments of $H^q_A$ do not satisfy 
polynomiality, even if the proton and neutron GPDs do. As we mentioned in the
Introduction, it is an outstanding theoretical challenge to
build a model of nuclear GPDs with the property of polynomiality.

DVCS observables are expressed in terms of the so-called Compton form  factors 
(CFFs), which are defined as nuclear GPDs convoluted with the 
corresponding hard scattering coefficients.
For spin-0 nuclei, to the leading order in $\alpha_s$, the only CFF reads
\begin{eqnarray}
{\cal H}_A(\xi_A,Q^2,t)&=&\sum e_q^2 \int^{1}_{-1} d x\, H_A^q(x,\xi_A,Q^2,t)
 \left(\frac{1}{x-\xi_A+i0}+\frac{1}{x+\xi_A-i0} \right) \nonumber\\
&=&\left(\frac{\xi_N}{\xi_A}\right)\sum e_q^2 \int^{1}_{-1}d x_N \Bigg[Z\left(H^{q/p}(x_N,\xi_N,Q^2,t) +\frac{t}{4 m_N^2}E^{q/p}(x_N,\xi_N,Q^2,t)
\right) \nonumber\\
&+&N\left(H^{q/n}(x_N,\xi_N,Q^2,t)+\frac{t}{4 m_N^2}E^{q/n}(x_N,\xi_N,Q^2,t)\right)
\Bigg] \nonumber\\
&\times&F_A(t)\left(\frac{1}{x_N-\xi_N+i0}+\frac{1}{x_N+\xi_N-i0} \right) \nonumber\\
&=&\left(\frac{\xi_N}{\xi_A}\right)\Bigg[Z
\left({\cal H}^p(\xi_N,Q^2,t)+\frac{t}{4 m_N^2}{\cal E}^p(\xi_N,Q^2,t) \right)
\nonumber\\
&+&N\left(
{\cal H}^n(\xi_N,Q^2,t)+\frac{t}{4 m_N^2}{\cal E}^n(\xi_N,Q^2,t) \right)
\Bigg] F_A(t) \,.
\label{eq:H_A_cal}
\end{eqnarray}
An important corollary of Eq.~(\ref{eq:H_A_cal}) is that 
${\cal H}_A$ scales as $A^2$.

In our analysis, 
for the nucleon CFFs ${\cal H}^{p,n}$ and ${\cal E}^{p,n}$, 
we used results of the dual 
parameterization of nucleon GPDs~\cite{Guzey:2006xi}, which gives a good description of the data on DVCS 
cross section and DVCS asymmetries on the proton target.
In the modeling of the nucleon GPD $E$, we took $J_u=J_d=0$.

For the nuclear form factor $F_A(t)$, for $^4$He, we used the result 
of~\cite{Frosch:1967pz}.
For other nuclei, we used
the parameterization
of nuclear charge density distributions~\cite{DeJager:1987qc} (see Appendix for details).

\section{Coherent 
and incoherent nuclear DVCS}
\label{sec:Interpolation}

\begin{figure}[t]
\begin{center}
\epsfig{file=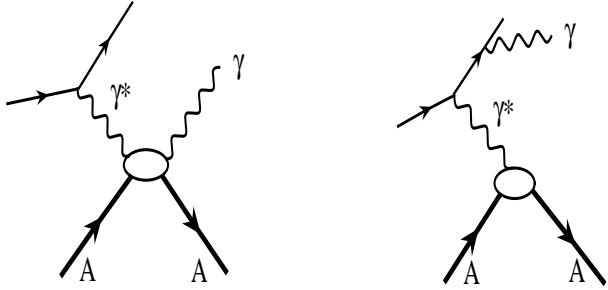,width=11cm,height=6cm}
\vskip -1cm
\caption{The coherent DVCS (left) and Bethe-Heitler (right)
 scattering amplitudes on a nucleus $A$.
Only one of two possible BH amplitudes is shown.}
\label{fig:coherent_fg}
\end{center}
\end{figure}

\begin{figure}[h]
\begin{center}
\epsfig{file=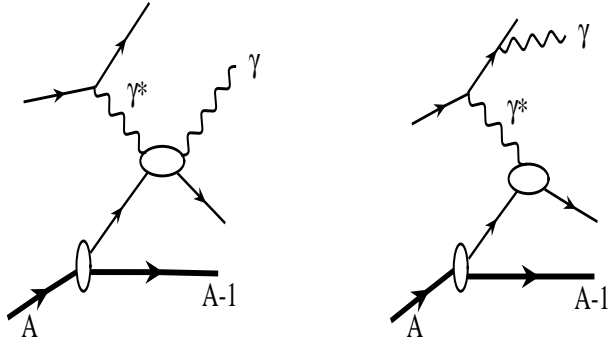,width=11cm,height=8cm}
\vskip -2cm
\caption{The incoherent DVCS and Bethe-Heitler scattering amplitudes.
The initial nucleus $A$ transforms into a final state containing $A-1$
spectator nucleons (free or bound) and an interacting nucleon.}
\label{fig:incoherent_fg}
\end{center}
\end{figure}

In the situation, when the recoiled nucleus is not detected,
measurements of DVCS observables with nuclear targets necessarily involve 
the coherent and incoherent contributions~\cite{Guzey:2003jh}. 
The former contribution corresponds to 
the case when the nuclear target stays intact, and it dominates at small 
values of the momentum transfer $t$.
 The latter 
contribution corresponds
to the case when the initial nucleus $A$ transforms into the system
of $A-1$ spectator nucleons (bound or free) and one interacting nucleon,
 and it dominates at large $t$.
In the approximation  of closure over the final nuclear
states, the exact structure of the final system of $A-1$ nucleons is not important.
The coherent DVCS and BH amplitudes (one of two possible attachments of the real photon 
to the lepton lines is shown)
 are presented in Fig.~\ref{fig:coherent_fg};
the incoherent DVCS and BH
amplitudes are shown in Fig.~\ref{fig:incoherent_fg}. 

In order to correctly sum the coherent and incoherent
contributions to the $e A \to e \gamma A$ cross section, 
let us schematically write the corresponding amplitude as, see e.g.~\cite{Frankfurt:2000jm},
\begin{equation}
{\cal A}(t)=\langle A^{\ast}| \sum_i^A J_i \,e^{i \vec{\Delta} \cdot \vec{r}_i} | A \rangle \,,
\label{eq:summation1}
\end{equation}
where $A^{\ast}$ represents the final state consisting of $A$ nucleons 
(coherently scattered nucleus or any product of the nuclear dissociation);
$J_i$ represents the operator corresponding to the interaction with the nucleon $i$
(one-particle operator);
the summation runs over all nucleons of the target; $\vec{\Delta}$ is the momentum
transfer.
Assuming that the states $|A^{\ast} \rangle$ form a complete set,
the cross section summed over the nuclear final states can be expressed in
the following form,
\begin{eqnarray}
\frac{d \sigma_A}{d t} &\propto& \sum_{A^{\ast}}\langle A | \sum_{j}^A J_j^{\dagger} \,e^{-i \vec{\Delta} \cdot \vec{r}_j}|A^{\ast}\rangle \langle A^{\ast}| \sum_i^A J_i\,e^{i \vec{\Delta} \cdot \vec{r}_i}
 | A \rangle=
 \langle A | \sum_{i,j}^A J_j^{\dagger}\, J_i \,e^{i \vec{\Delta} \cdot (\vec{r}_i- \vec{r}_j)}
| A \rangle \nonumber\\
&=& \langle A | \sum_{i\neq j}^A J_j^{\dagger}\, J_i \,e^{i \vec{\Delta} \cdot (\vec{r}_i- \vec{r}_j)}| A \rangle+
\langle A | \sum_{i}^A J_i^{\dagger}\, J_i | A \rangle \nonumber\\
& \approx &A(A-1) \, \langle A |J_N^{\dagger}\, J_N \,e^{i \vec{\Delta} \cdot (\vec{r}_i- \vec{r}_j)}| A \rangle+A\,\langle N |J_N^{\dagger}\, J_N | N \rangle
\nonumber\\
& \propto & A(A-1)\, F_A^2(t^{\prime}) \, \frac{d \tilde{\sigma}_N}{dt}+
A \,\frac{d \sigma_N}{dt}
\,,
\label{eq:summation2}
\end{eqnarray}
where $d \tilde{\sigma}_N/dt$ is the scattering cross section on the bound 
nucleon; $d \sigma_N/dt$ corresponds to the quasi-free nucleon;
$t^{\prime}=A/(A-1)\,t$~\cite{Frankfurt:2000jm}.
For the sake of the argument, we did not distinguish between protons
and neutrons. 
Adopting the HERMES terminology, we shall call the
first term in the last line of Eq.~(\ref{eq:summation2})
{\it coherent-enriched}~\cite{Ellinghaus:2007dw}.
The second term is the incoherent contribution.

The dependence of the coherent-enriched contribution 
on $t$ is steep
and is governed by the nuclear form factor squared $F_A^2(t^{\prime})$.
Therefore, this contribution dominates the nuclear cross section at small
$t$.
The $t$-dependence of the incoherent 
contribution is much slower and is determined by the $t$-dependence of the 
cross section on quasi-free nucleons $ d \sigma_N/dt$. 
While this contribution is present at all $t$, it dominates the 
nuclear cross section at large $t$.

Besides the $t$-dependence, the coherent-enriched and incoherent contributions
have different $A$-dependences. 
The coherent-enriched contribution scales as $A(A-1)$; the incoherent 
contribution scales as $A$.

Let us now consider 
the case, when the recoiled nucleus is intact.
 In this case, $|A^{\ast} \rangle=|A \rangle$ in
Eq.~(\ref{eq:summation1}), and
the expression for the $e A \to e \gamma A$ cross section becomes
\begin{equation}
\frac{d \sigma_A}{d t} =A^2 F_A^2(t) \, \frac{d \tilde{\sigma}_N}{dt}
\,.
\label{eq:summation3}
\end{equation}
In Eq.~(\ref{eq:summation3}), $d \sigma_A/d t$ is the 
genuine coherent nuclear scattering cross section, which scales as $A^2$ and whose 
$t$-dependence is steep and is given by the nuclear
form factor squared $F_A^2(t)$.

Using Eq.~(\ref{eq:summation2}),
the full-fledged differential cross section for the $e A \to e \gamma A$ reaction~\cite{Belitsky:2001ns} can be written 
as a sum of the coherent-enriched and incoherent  contributions,
\begin{equation}
\frac{d \sigma_A}{dx_A d y d t d \phi}=\frac{\alpha^3 x_A y}{8 \pi Q^2 \sqrt{1+\epsilon^2}} \left(\frac{A-1}{A}\left|\frac{{\cal T}_A(x_A,y)}{e^3}\right|^2
+\left(\frac{x_B}{x_A}\right)^2
\sum_{i=1}^{A}\left|\frac{{\cal T}_i(x_B,y)}{e^3}\right|^2
\right) \,,
\label{eq:cs1}
\end{equation}
where ${\cal T}_A$ is the amplitude for the coherent $e A \to e \gamma A$
scattering; ${\cal T}_i$ are the amplitudes for quasi-free
incoherent $e A \to e \gamma A$ scattering;
the $(A-1)/A$ factor  originates from Eq.~(\ref{eq:summation2});
the $(x_B/x_A)^2$ factor is required for the incoherent
contribution not to depend on $A$; 
$\phi$ is the angle between the lepton scattering and the production planes.

It is important to note that the prefactor $(A-1)/A$ corresponds to the DVCS amplitude squared and to the interference between the DVCS and Bethe-Heitler (BH) amplitudes.
For the BH amplitude squared, $(A-1)/A$ should be replaced by $(Z-1)/Z$.

In Eq.~(\ref{eq:cs1}), in the laboratory frame,
\begin{equation}
y=\frac{\nu}{E} \,,
\quad \epsilon=2 \frac{x_A M_A}{Q}=2 \frac{x_B m_N}{Q}
\,,
\label{eq:yA}
\end{equation}
where $E$ is the energy (momentum) of the incoming lepton.
Note that
the variables $y$, $\epsilon$ and $t$ are the same for nuclear and nucleon targets.

For the comparison with the  free nucleon case and with 
experiments,
it is convenient to express $\sigma_A$ as a function of $x_B$,
\begin{equation}
\frac{d \sigma_A}{dx_B d y d t d \phi}=\frac{\alpha^3 x_B y}{8 \pi Q^2 \sqrt{1+\epsilon^2}} \left(\frac{A-1}{A^3}\left|\frac{{\cal T}_A(x_A,y)}{e^3}\right|^2
+
\sum_{i=1}^{A}\left|\frac{{\cal T}_i(x_B,y)}{e^3}\right|^2
\right) \,.
\label{eq:cs2}
\end{equation}
For the BH amplitude squared, $(A-1)/A^3$ should be replaced by $(Z-1)/(Z A^2)$.

For illustration, let us consider the DVCS contribution to 
Eq.~(\ref{eq:cs2}).
In this case, $|{\cal T}_A|^2 \propto |{\cal H}_A|^2$, which scales as
$[A^2 F_A(t^{\prime})]^2$, see Eq.~(\ref{eq:H_A_cal}).
 Therefore, the first term in Eq.~(\ref{eq:cs2}) 
behaves as $A^2 F_A^2(t^{\prime})$.
The second term has the $t$-dependence determined by
the nucleon GPDs and scales as $A$.

In the situation, when the recoiled nucleus is detected,
the $e A \to e \gamma A$ cross section is purely coherent, and it reads
\begin{equation}
\frac{d \sigma_A}{dx_B d y d t d \phi}=\frac{\alpha^3 x_B y}{8 \pi Q^2 \sqrt{1+\epsilon^2}} \frac{1}{A^2}\left|\frac{{\cal T}_A(x_A,y)}{e^3}\right|^2
 \,.
\label{eq:cs3}
\end{equation}

\section{Nuclear DVCS asymmetries}
\label{sec:nucl_asymmetry}

In this section, as an example of DVCS asymmetries, 
we consider the beam-spin nuclear DVCS asymmetry
in the presence of the coherent and incoherent contributions,
with an emphasis on the neutron contribution. 
We make predictions relevant for the HERMES and JLab kinematics.

\subsection{Coherent and incoherent contributions to DVCS asymmetries}
\label{subsec:asymmetry}

Expressions for nuclear DVCS asymmetries can be readily obtained
from Eqs.~(\ref{eq:cs2}) and (\ref{eq:cs3}).
In this work, we consider the beam-spin asymmetry, $A_{LU}$, which
is measured with the longitudinally-polarized lepton beam and the unpolarized 
target.

The nuclear and nucleon
amplitudes squared in Eqs.~(\ref{eq:cs2}) and (\ref{eq:cs3})
receive contributions from
the DVCS and Bethe-Heitler (BH) scattering amplitudes and their interference,
\begin{equation}
|{\cal T}|^2=|{\cal T}_{{\rm DVCS}}|^2+|{\cal T}_{{\rm BH}}|^2+{\cal I} \,,
\label{eq:T}
\end{equation}
where ${\cal I}={\cal T}_{{\rm DVCS}}^{\ast}{\cal T}_{{\rm BH}}
+{\cal T}_{{\rm BH}}^{\ast}{\cal T}_{{\rm DVCS}}$.

The expression for the nuclear DVCS beam-spin asymmetry  reads~\cite{Belitsky:2001ns}
\begin{equation}
A_{LU}^A(\phi)=\frac{\Delta {\cal I}}
{|{\cal T}_{{\rm BH}}|^2+{\cal I}+|{\cal T}_{{\rm DVCS}}|^2} \,,
\label{eq:as1}
\end{equation}
where  
$\Delta {\cal I}=1/2({\cal I}^{\lambda=1}
-{\cal I}^{\lambda=-1})$ with
$\lambda$ the helicity of
the incoming lepton; all other contributions correspond to the unpolarized
beam.

In the situation corresponding to Eq.~(\ref{eq:cs2}),
each term in Eq.~(\ref{eq:as1}) contains the coherent-enriched   and 
incoherent  contributions,
\begin{eqnarray}
{\cal I} &=& \frac{A-1}{A^3}\, {\cal I}^A+ 
Z\, {\cal I}^p+N \,{\cal I}^n
\,, \nonumber\\
|{\cal T}_{{\rm BH}}|^2 &=&\frac{Z-1}{Z A^2} \,  |{\cal T}_{{\rm BH}}^A|^2
+ Z\, |{\cal T}_{{\rm BH}}^p|^2
+N \,|{\cal T}_{{\rm BH}}^n|^2 
\,, \nonumber\\
|{\cal T}_{{\rm DVCS}}|^2 &=&\frac{A-1}{A^3}\, |{\cal T}_{{\rm DVCS}}^{A}|^2+ 
Z\, |{\cal T}_{{\rm DVCS}}^{p}|^2+N \,|{\cal T}_{{\rm DVCS}}^{n}|^2 
\,.
\label{eq:as2}
\end{eqnarray}
Expressions for the free nucleon contributions ${\cal I}^{p,n}$,
$|{\cal T}_{{\rm BH}}^{p,n}|^2$ and 
$|{\cal T}_{{\rm DVCS}}^{p,n}|^2$ in terms of $\cos \phi$ and
$\sin \phi$-harmonics are derived in~\cite{Belitsky:2001ns}.
As a model of the nucleon GPDs, we used the results of the dual
parameterization of nucleon GPDs with $J_u=J_d=0$~\cite{Guzey:2006xi}.

Expressions for ${\cal I}^{A}$,
$|{\cal T}_{{\rm BH}}^A|^2$ and 
$|{\cal T}_{{\rm DVCS}}^{A}|^2$ 
for spin-0 zero nuclei are the same as for the pion~\cite{Belitsky:2000vk},
after the replacement of the pion charge form factor by the nuclear one
evaluated at $t^{\prime}=A/(A-1)t$.

In the case of the purely coherent scattering corresponding to Eq.~(\ref{eq:cs3}),
the terms in Eq.~(\ref{eq:as2}) should be replaced by the following expressions,
\begin{eqnarray}
{\cal I} &=& \frac{1}{A^2}\, {\cal I}^A
\,, \nonumber\\
|{\cal T}_{{\rm BH}}|^2 &=&\frac{1}{A^2} \,  |{\cal T}_{{\rm BH}}^A|^2
\,, \nonumber\\
|{\cal T}_{{\rm DVCS}}|^2 &=&\frac{1}{A^2}\, |{\cal T}_{{\rm DVCS}}^{A}|^2
\,.
\label{eq:as3}
\end{eqnarray}
In the purely coherent case, the nuclear form factor is evaluated at 
the momentum transfer $t$.

Using Eqs.~(\ref{eq:as1}), (\ref{eq:as2}) and (\ref{eq:as3}), one can 
qualitatively estimate the behavior of $A_{LU}^A(\phi)$ as a function of $A$ 
and $Z$. Provided the $|{\cal T}_{{\rm BH}}|^2$-term dominates the unpolarized
cross section, the coherent-enriched contribution to $A_{LU}^A(\phi)$ scales
as $(A-1)/(Z-1)$. The purely coherent $A_{LU}^A(\phi)$ scales as $A/Z$.

All expressions used in Eqs.~(\ref{eq:as2}) and (\ref{eq:as3}) are collected in
 Appendix.

\subsection{Nuclear DVCS beam-spin asymmetry $A_{LU}$ in HERMES kinematics}
\label{subsec:ALU_Hermes}

In the measurement of nuclear DVCS at HERMES,
the recoiled nucleus is not detected, but  reconstructed using the missing
mass technique~\cite{Ellinghaus:2002zw,Ellinghaus:2007dw}. This corresponds to the situation, when
one sums over all final nuclear states. This means that the nuclear beam-spin
DVCS asymmetry, $A_{LU}^A$, receives contribution from the coherent-enriched and 
incoherent terms and is given by Eqs.~(\ref{eq:as1}) and (\ref{eq:as2}).

\begin{figure}[t]
\begin{center}
\epsfig{file=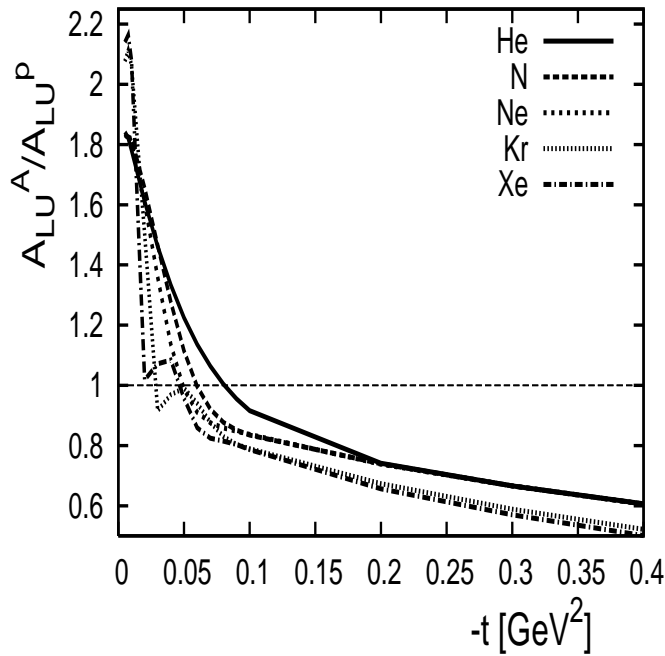,width=9cm,height=9cm}
\caption{The ratio of the nuclear to free proton beam-spin DVCS asymmetries,
$A_{LU}^A(\phi)/A_{LU}^p(\phi)$, as a function of the momentum transfer $t$ for 
He, N, Ne, Kr and Xe nuclei. 
The calculation is done at $x_B=0.065$, $Q^2=1.7$ GeV$^2$~\cite{Ellinghaus:2007dw}
and $\phi=90^{\circ}$.}
\label{fig:nuclear_beam_spin_tdep}
\end{center}
\end{figure}

We quantify our numerical predictions for $A_{LU}^A$ by considering the ratio
of the nuclear to the free proton asymmetries,
$A_{LU}^A(\phi)/A_{LU}^p(\phi)$. This ratio is presented in 
Fig.~\ref{fig:nuclear_beam_spin_tdep} as a function of $t$ at an average
HERMES kinematic point $x_B=0.065$ and $Q^2=1.7$ GeV$^2$~\cite{Ellinghaus:2007dw}.
The asymmetries are evaluated at $\phi=90^{\circ}$. 
 Different curves correspond to different nuclei: $^{4}$He,
$^{14}$N, $^{20}$Ne, $^{84}$Kr and $^{131}$Xe.

At small values of $t$, when the nuclear asymmetries (cross sections) are dominated
by the coherent-enriched contribution, $A_{LU}^A/(\phi)A_{LU}^p(\phi)=1.8-2.2$,
which is consistent with the previous analyses~\cite{Guzey:2003jh,Kirchner:2003wt}.
The enhancement of $A_{LU}^A(\phi)/A_{LU}^p(\phi)$ above unity is the combinatoric effect:
Since the interference between the Bethe-Heitler and the DVCS amplitudes 
scales as $Z (A-1)$ and the Bethe-Heitler amplitude squared scales as $Z(Z-1)$, 
$A_{LU}^A(\phi)$ scales as $(A-1)/(Z-1)$.

At large values of $t$, when the nuclear form factor eliminates the
coherent-enriched term, $A_{LU}^A(\phi)$ is given by 
the incoherent contribution, and $A_{LU}^A(\phi)/A_{LU}^p(\phi) < 1$.

The fact that  $A_{LU}^A(\phi)/A_{LU}^p(\phi) < 1$ is a result of the neutron contribution to $A_{LU}^A(\phi)$, see Eq.~(\ref{eq:as2}). 
First (this is effect is largest),
the neutron contribution decreases the numerator of
$A_{LU}^A(\phi)$, since $F_{1n} <0$, while $F_{1p} >0$.
Second, the positive neutron contributions 
$|{\cal T}_{{\rm BH}}^n|^2+{\cal I}^n$ 
(somewhat
suppressed by the neutron electromagnetic form factors compared to the proton 
contribution)
 and $|{\cal T}_{{\rm DVCS}}^n|^2$ (similar to the proton contribution)
increase the denominator of $A_{LU}^A(\phi)$.
The decrease of the numerator of $A_{LU}^A(\phi)$ and the increase of the denominator
work together to reduce $A_{LU}^A(\phi)/A_{LU}^p(\phi)$ significantly below
unity at large $t$.

Note that our present finding that $A_{LU}^A(\phi)/A_{LU}^p(\phi) < 1$ 
at large $t$
does not contradict the original analysis~\cite{Guzey:2003jh}. 
In that work, is was predicted that $A_{LU}^A(\phi)/A_{LU}^p(\phi) \to 1$ as
$t$ becomes large, if the neutron contribution to the nuclear asymmetry is
neglected. In the present work, we went beyond this approximation and found that
the neutron contribution is not negligible and leads 
to $A_{LU}^A(\phi)/A_{LU}^p(\phi) < 1$.
Therefore, studies of the incoherent contribution to nuclear DVCS
asymmetries is a sensitive tool to constrain neutron GPDs.
The CLAS collaboration at Jefferson Lab explored this 
possibility using the deuterium target~\cite{Mazouz:2007vj}.

\begin{figure}[t]
\begin{center}
\epsfig{file=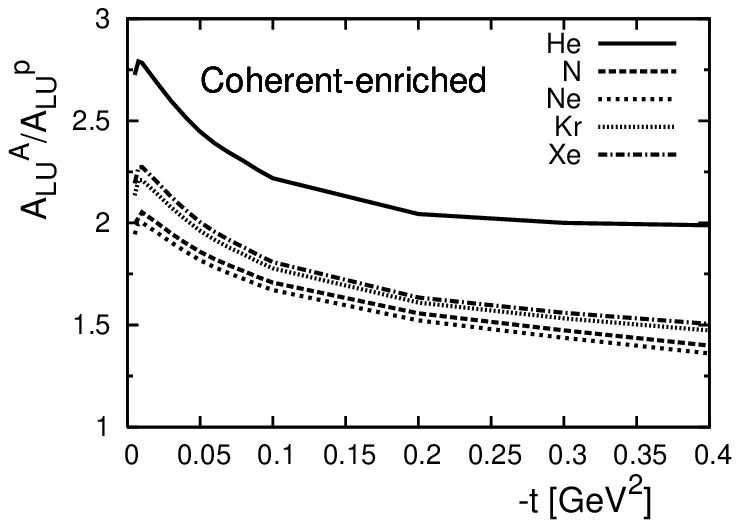,width=8cm,height=8cm}
\epsfig{file=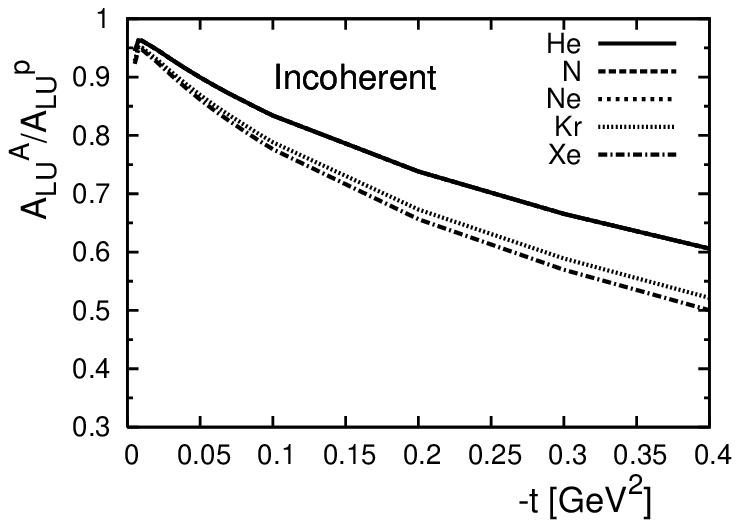,width=8cm,height=8cm}
\caption{The ratio of the nuclear to free proton beam-spin DVCS asymmetries,
$A_{LU}^A(\phi)/A_{LU}^p(\phi)$, as a function of the momentum transfer $t$ for 
He, N, Ne, Kr and Xe nuclei. 
The calculation is done at $x_B=0.065$, $Q^2=1.7$ GeV$^2$~\cite{Ellinghaus:2007dw}
and $\phi=90^{\circ}$. 
The left panel corresponds to the coherent-enriched contribution; the right panel --
to the incoherent contribution.
}
\label{fig:nuclear_beam_spin_tdep2}
\end{center}
\end{figure}

Note also that the neutron GPDs enter the model of nuclear GPDs, see
Eq.~(\ref{eq:model1}). Hence, 
nuclear DVCS observables in the coherent regime also
 provide certain constraints for  the neutron GPDs, albeit those
constraints are less stringent and more model-dependent compared to the
incoherent regime.

By studying the $t$-dependence of the nuclear DVCS cross section, the HERMES
analysis separated the coherent-enriched and incoherent
contributions to $A_{LU}^A(\phi)$. 
Our predictions for these two contributions are presented separately in
Fig.~\ref{fig:nuclear_beam_spin_tdep2}.
The left panel corresponds to the coherent-enriched
contribution to $A_{LU}^A(\phi)$, which was calculated keeping only 
first terms in Eq.~(\ref{eq:as2}). The right panel corresponds to the
incoherent contribution calculated using last two terms in 
Eq.~(\ref{eq:as2}).

In the left panel of Fig.~\ref{fig:nuclear_beam_spin_tdep2},
the curve for $^4$He lies above the curves for other nuclei because the 
coherent-enriched contribution scales $(A-1)/(Z-1)$.

In the right panel of Fig.~\ref{fig:nuclear_beam_spin_tdep2},
the ratio $A_{LU}^A(\phi)/A_{LU}^p(\phi)$ at small $t$ is close to unity
because the neutron contribution is suppressed by the small value of
the neutron Dirac form factor
$F_{1n}(t)$. As $|F_{1n}(t)|$ increases with increasing $|t|$,
the ratio $A_{LU}^A(\phi)/A_{LU}^p(\phi)$
begins to progressively deviate from unity.

\begin{figure}[t]
\begin{center}
\epsfig{file=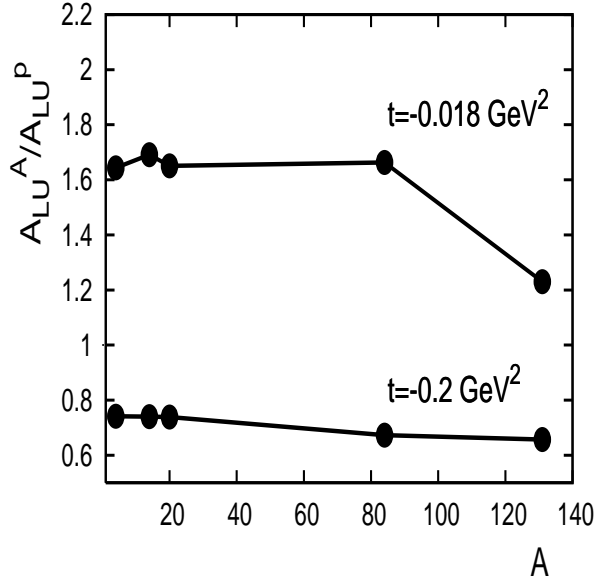,width=8cm,height=8cm}
\caption{The ratio of the nuclear to free proton beam-spin DVCS asymmetries,
$A_{LU}^A(\phi)/A_{LU}^p(\phi)$, as a function of $A$.
The calculation is done at $x_B=0.065$, $Q^2=1.7$ GeV$^2$
and $\phi=90^{\circ}$.
}
\label{fig:nuclear_beam_spin_Adep}
\end{center}
\end{figure}

Taking different $t$-slices of Fig.~\ref{fig:nuclear_beam_spin_tdep}, 
we can study the $A$-dependence of $A_{LU}^A(\phi)$. 
Figure~\ref{fig:nuclear_beam_spin_Adep} presents $A_{LU}^A(\phi)/A_{LU}^p(\phi)$
as a function of $A$ at $t=-0.018$ GeV$^2$ (upper set of points) and $t=-0.2$  GeV$^2$
(lower set of points). These two values of $t$ correspond to the average
HERMES values~\cite{Ellinghaus:2007dw}.

The interpretation of Fig.~\ref{fig:nuclear_beam_spin_Adep} is the same as for
Fig.~\ref{fig:nuclear_beam_spin_tdep}. At small values of $t$, the coherent-enriched
contribution dominates and $A_{LU}^A(\phi)/A_{LU}^p(\phi)>1$ due to the fact that
$A_{LU}^A(\phi)$ scales roughly as $(A-1)/(Z-1)$.
At large $t$, where only the incoherent contribution matters,
$A_{LU}^A(\phi)/A_{LU}^p(\phi)< 1$ due to the neutron contribution (see the discussion 
above).

Results presented in Fig.~\ref{fig:nuclear_beam_spin_Adep} should be compared
to the results of the HERMES analysis~\cite{Ellinghaus:2007dw}. 
At $t=-0.018$ GeV$^2$, the agreement between our calculations 
(the upper set of points) and the HERMES data is excellent. For the
nuclei of $^{4}$He, $^{14}$N, $^{20}$Ne and $^{84}$Kr, $A_{LU}^A(\phi)/A_{LU}^p(\phi) \approx 1.65$.
 For the nucleus 
of $^{131}$Xe,  $A_{LU}^A(\phi)/A_{LU}^p(\phi)=1.23$, which is smaller than for other
lighter nuclei because  of the reduction of the coherent-enriched contribution
by the nuclear form factor.

At $t=-0.2$ GeV$^2$, we predict that $A_{LU}^A(\phi)/A_{LU}^p(\phi) = 0.66-0.74$,
depending of the target nucleus.
The experimental uncertainties of the HERMES data are too large and, in general, 
do not exclude the deviation of $A_{LU}^A(\phi)/A_{LU}^p(\phi)$ from unity, as we predict.
 
\subsection{Nuclear DVCS beam-spin asymmetry $A_{LU}$ in Jefferson Lab kinematics}
\label{subsec:ALU_CLAS}

There exists an exciting possibility to study purely
coherent nuclear DVCS at Jefferson Lab using the 
BoNuS recoil detector. In particular, an experiment to study
$A_{LU}$ in coherent and incoherent DVCS on $^4$He has been proposed.
Main advantages of the proposed experiment compared to HERMES  is 
exclusivity of the measurement, 
which will allow to measure the purely coherent
DVCS, and small projected errors due to high statistics, which will enable one to unambiguously determine the 
magnitude of  $A_{LU}$ in the coherent and incoherent regimes.
In addition, the proposed experiment might shed some light on the question of modifications
of nucleon GPDs in nuclear medium.

\begin{figure}[t]
\begin{center}
\epsfig{file=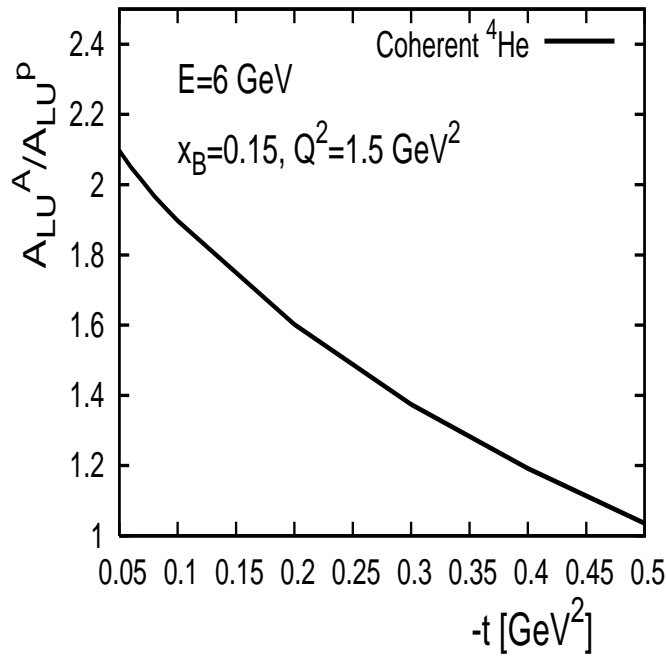,width=9cm,height=9cm}
\caption{The ratio of the coherent $^4$He to free proton beam-spin DVCS asymmetries,
$A_{LU}^A(\phi)/A_{LU}^p(\phi)$, as a function of the momentum transfer $t$. 
The calculation corresponds to JLab kinematics, $E=6$ GeV, $x_B=0.15$, $Q^2=1.5$ 
GeV$^2$,
and was performed at $\phi=90^{\circ}$.}
\label{fig:nuclear_beam_spin_tdep_JLab}
\end{center}
\end{figure}
 
Figure~\ref{fig:nuclear_beam_spin_tdep_JLab} presents our predictions for the 
ratio of the coherent $^4$He to free proton beam-spin DVCS asymmetries,
$A_{LU}^A(\phi)/A_{LU}^p(\phi)$, as a function of the momentum transfer $t$. 
The calculation corresponds to a typical point in the  
current  JLab kinematics: $E=6$ GeV, $x_B=0.15$ and $Q^2=1.5$ GeV$^2$.
The ratio of the asymmetries is evaluated at $\phi=90^{\circ}$.

The behavior of $A_{LU}^A(\phi)/A_{LU}^p(\phi)$ presented in 
Fig.~\ref{fig:nuclear_beam_spin_tdep_JLab} is similar to that in
the left-hand side panel of Fig.~\ref{fig:nuclear_beam_spin_tdep2}.
Since the purely coherent $A_{LU}^A(\phi)$ scales as $A/Z$, 
while the coherent-enriched contribution to $A_{LU}^A(\phi)$
scales as $(A-1)/(Z-1)$, the curve in 
Fig.~\ref{fig:nuclear_beam_spin_tdep_JLab}
lies lower than the corresponding curve in 
the left-hand side of Fig.~\ref{fig:nuclear_beam_spin_tdep2}.

In relation to incoherent DVCS on $^4$He, the proposed experiment at Jefferson Lab
will measure the $e ^4{\rm He} \to e p X$ reaction, i.e.~DVCS on a quasi-free proton.
In this case, the neutron contribution is absent and the ratio 
$A_{LU}^A(\phi)/A_{LU}^p(\phi)=1$, provided the bound proton in $^4$He
is not modified.
Therefore, this measurement will probe modifications of proton GPDs in $^4$He.

\section{Conclusions and discussion}
\label{sec:SD}

Using a simple model for nuclear GPDs, we studied 
the role of the neutron 
contribution to nuclear DVCS observables. As an example,
we used the beam-spin asymmetry $A_{LU}^A$ measured in coherent and 
incoherent DVCS on a wide range of nuclear targets.
In our analysis, we considered the $^4$He, $^{14}$N, $^{20}$Ne, $^{84}$Kr and
$^{131}$Xe nuclei in the HERMES kinematics and the  $^4$He nucleus
in the JLab kinematics.

We found that at small values of the momentum transfer $t$, 
$A_{LU}^A$ is dominated by the  coherent-enriched contribution, which scales as
$(A-1)/(Z-1)$. 
This enhances the nuclear $A_{LU}^A$ compared to the free proton
$A_{LU}^p$, $A_{LU}^A(\phi)/A_{LU}^p(\phi)=1.8-2.2$,
in accord with the earlier predictions~\cite{Guzey:2003jh,Kirchner:2003wt}.

On the other hand,
at large values
of $t$, when the nuclear asymmetry is dominated by the incoherent
contribution,  $A_{LU}^A(\phi)/A_{LU}^p(\phi)$ is significantly smaller than unity:
$A_{LU}^A(\phi)/A_{LU}^p(\phi)=0.66-0.74$, depending on the target nucleus.
This deviation of $A_{LU}^A(\phi)/A_{LU}^p(\phi)$ from unity is a result of the
neutron contribution: 
The negative neutron contribution ($F_{1n} <0$) decreases the numerator
of $A_{LU}^A$ and, at the same time, the positive 
neutron contribution  
$|{\cal T}_{{\rm BH}}^n|^2+{\cal I}^n+|{\cal T}_{{\rm DVCS}}^n|^2$
increases the denominator of $A_{LU}^A$.
Since the effect of the deviation of $A_{LU}^A(\phi)/A_{LU}^p(\phi)$ from unity is so 
sizable, incoherent DVCS on nuclei gives a possibility to constain neutron GPDs.

In this work, we considered  one kind of DVCS observables, namely,
the beam-spin asymmetry. We expect that for other DVCS asymmetries, such 
as e.g.~for the beam-charge DVCS asymmetry, the ratio of the nuclear to the free
proton asymmetries will be qualitatively similar to 
 $A_{LU}^A(\phi)/A_{LU}^p(\phi)$, see~\cite{Guzey:2003jh}.

All results presented in this work, data grids for the 
dual parameterization of the nucleon GPDs and FORTRAN codes for various DVCS
asymmetries measured in DVCS on nucleons and nuclei
can be found and downloaded from the author's webpage
{\tt http://www.jlab.org/\~{}vguzey}.

\acknowledgments

We would like to thank H.~Egiyan, F.X.~Girod, H.~Guler, K.~Hafidi,
D.~Hash  and M.~Strikman for 
useful and encouraging discussions.

{\bf Notice}: Authored by Jefferson Science Associates, LLC under U.S. DOE Contract No. DE-AC05-06OR23177. The U.S. Government retains a non-exclusive, paid-up, irrevocable, world-wide license to publish or reproduce this manuscript for U.S. Government purposes.

\appendix*
\section{Input for calculation of DVCS asymmetries}

In this Appendix, we collect all expressions used in our numerical
analysis of the nuclear and proton DVCS beam-spin asymmetries, 
see Eqs.~(\ref{eq:as1}), (\ref{eq:as2}) and (\ref{eq:as3}).

The interference, Bethe-Heitler and DVCS terms, which enter 
Eqs.~(\ref{eq:as1}), (\ref{eq:as2}) and (\ref{eq:as3}),
 read~\cite{Belitsky:2001ns}
\begin{eqnarray}
{\cal I}&=&\frac{\pm e^6}{x y^3 t {\cal P}_1(\phi){\cal P}_2(\phi)}
\Big(c_{0,{\rm unp}}^{\cal I}+ 
c_{1,{\rm unp}}^{\cal I} \cos (\phi)+s_{1,{\rm unp}}^{\cal I} \sin (\phi) 
 \Big) \,, \nonumber\\
|{\cal T}_{{\rm BH}}|^2&=&\frac{e^6}{x^2 y^2 (1+\epsilon^2)^2 t {\cal P}_1(\phi){\cal P}_2(\phi)}
\Big(c_{0,{\rm unp}}^{\rm BH}+c_{1,{\rm unp}}^{\rm BH} \cos (\phi)+c_{2,{\rm unp}}^{\rm BH} \cos (2\phi) 
 \Big) \,, \nonumber\\
|{\cal T}_{{\rm DVCS}}|^2&=&\frac{e^6}{y^2 Q^2}\,c_{0,{\rm unp}}^{\rm DVCS} \,,
\label{eq:a1}
\end{eqnarray}
where we have kept only twist-two terms and neglected the gluon GPDs.
In Eq.~(\ref{eq:a1}), ${\cal P}_1(\phi)$ and ${\cal P}_2(\phi)$
are lepton propagators; the plus-sign in front of the interference term
corresponds to electrons, while the minus-sign is for positrons;
$\phi$ is the angle between the lepton and production planes;
the coefficients $c_{0,1,{\rm unp}}^I$, $s_{1,{\rm unp}}^I$, $c_{0,1,2}^{{\rm BH}}$ 
and $c_{0,{\rm unp}}^{{\rm DVCS}}$ are called harmonics.

When Eq.~(\ref{eq:a1}) is applied to the 
coherent-enriched 
contribution, it should be  evaluated with $x=x_A$
and the corresponding nuclear harmonics (see below).
When Eq.~(\ref{eq:a1}) is used to calculate the incoherent
contribution, it should be  evaluated 
with with $x=x_B$ and with the free proton and neutron 
harmonics (see below).

\subsection{Nuclear part}

Expressions for the $\cos \phi$ and $\sin \phi$-harmonics
of a spin-0 zero nucleus are the same as for the pion 
case~\cite{Belitsky:2000vk} after the replacement of the pion GPD and 
the electric form factor by their nuclear counterparts
(one has also divide the pion harmonics involving GPDs by the factor of $x$ 
due to a different normalization of the interference and DVCS terms used 
in~\cite{Belitsky:2000vk}).
The required harmonics read
\begin{eqnarray}
c_{0,{\rm unp}}^{\cal I}& = & -8 \frac{t}{Q^2} (2-y) \left[(2-x_A)(1-y)-(1-x_A)(2-y)^2\left(1-\frac{t_{{\rm min}}}{t}\right) \right] ZF_A(t)\, \Re e {\cal H}_A
\,, \nonumber\\
c_{1,{\rm unp}}^{\cal I}& = & -8 \,K\, (2-2y+y^2)\, Z F_A(t)\, \Re e {\cal H}_A\,, \nonumber\\
s_{1,{\rm unp}}^{\cal I}& = & 8\, K\,\lambda\, y\, (2-y)\,ZF_A(t)\, \Im m {\cal H}_A \,, \nonumber\\
c_{0,{\rm unp}}^{{\rm BH}} & = & \Bigg\{\left((2-y)^2+y^2(1+\epsilon^2)^2 \right)
\left[4\,x_A^2 \frac{M_A^2}{t} +4(1-x_A)+(4x_A+\epsilon^2)\frac{t}{Q^2} \right]
+ 32\,x_A^2 K^2 \frac{M_A^2}{t}
\nonumber\\
&+&2\,\epsilon^2 \left[4 (1-y)(3+2 \epsilon^2)+y^2(2-\epsilon^4) \right]
-4 x_A^2(2-y)^2 (2+\epsilon^2) \frac{t}{Q^2} \Bigg\} Z^2 F_A^2(t) \,,
\nonumber\\
c_{1,{\rm unp}}^{{\rm BH}} & = & -8\,K\,(2-y) 
\left(2\,x_A+\epsilon^2-4\,x_A^2\frac{M_A^2}{t}\right)\,Z^2 F_A^2(t) \,,
\nonumber\\
c_{2,{\rm unp}}^{{\rm BH}} & = & 32\,K^2\,x_A^2 \frac{M_A^2}{t}\,Z^2F_A^2(t) \,,
\nonumber\\
c_{0,{\rm unp}}^{{\rm DVCS}} & = & 2(2-2y+y^2)\, |{\cal H}_A|^2 \,,
\label{eq:a2}
\end{eqnarray}
where $K$ is the so-called kinematic $K$-factor~\cite{Belitsky:2001ns};
$\lambda$ is the incoming lepton helicity.

The nuclear form factor $F_A$ entering Eq.~(\ref{eq:a2}) is 
evaluated at $t^{\prime}=A/(A-1)t$ for the coherent-enriched contribution and
at $t$ for the purely coherent case.

For $^4$He, the nuclear form factor is parameterized as
\begin{equation}
F_A(t)=\left(1-(a^2 t)^6\right) e^{-b^2|t|} \,,
\end{equation}
where $a=0.316$ fm and $b=0.681$ fm~\cite{Frosch:1967pz}.

For other nuclei used in this paper, the nuclear form factor is defined as
\begin{equation}
F_A(t)=4 \pi \int_0^{\infty} d r r \frac{\sin(\sqrt{|t|}r)}{\sqrt{|t|}}
 \rho_A(r) \,,
\end{equation}
where $\rho_A(r)$ is the nuclear charge density distribution taken in the following form~\cite{DeJager:1987qc}.

Nitrogen ($A=14$, $Z=7$):
\begin{eqnarray}
w &=&-0.18 \,, \quad
z=0.505 \,, \nonumber\\
c&=&2.57 \,, \quad
\rho_0=0.0127908 \,, \nonumber\\
\rho_A(r)&=&\rho_0 \frac{1+w \frac{r^2}{c^2}}{1+e^{(r-c)/z}} \,.
\end{eqnarray}

Neon ($A=20$, $Z=10$):
\begin{eqnarray}
z&=&0.571 \,, \quad
c=2.805 \,, \nonumber\\
\rho_0 &=& 0.00767524 \,, \nonumber\\
\rho_A(r)&=&\rho_0 \frac{1}{1+e^{(r-c)/z}} \,.
\end{eqnarray}

Krypton ($A=84$, $Z=36$):
\begin{eqnarray}
z&=&0.496 \,, \quad
c=4.83 \,, \nonumber\\
\rho_0 &=& 0.00191897 \,, \nonumber\\
\rho_A(r)&=&\rho_0 \frac{1}{1+e^{(r-c)/z}} \,.
\end{eqnarray}

Xenon ($A=131$, $Z=54$):
\begin{eqnarray}
w &=&0.3749 \,, \quad
z=2.6776 \,, \nonumber\\
c&=&5.3376 \,, \quad
\rho_0=0.00112617 \,, \nonumber\\
\rho_A(r)&=&\rho_0 \frac{1+w \frac{r^2}{c^2}}{1+e^{(r^2-c^2)/z^2}} \,.
\end{eqnarray}

\subsection{Proton part}

Expressions for the required $\cos \phi$ and $\sin \phi$-harmonics for the 
proton target are derived in~\cite{Belitsky:2001ns}
\begin{eqnarray}
c_{0,{\rm unp}}^{\cal I}& = & -8\,(2-y) \Re e \left[\frac{(2-y)^2}{1-y} K^2 C^{\cal I}_{{\rm unp}} +\frac{t}{Q^2}(1-y)(2-x_B)(C^{\cal I}_{{\rm unp}}+\Delta C^{\cal I}_{{\rm unp}}) \right] 
\,, \nonumber\\
c_{1,{\rm unp}}^{\cal I}& = & -8 \,K\, (2-2y+y^2)\, \, \Re e\, C^{\cal I}_{{\rm unp}} \,, \nonumber\\
s_{1,{\rm unp}}^{\cal I}& = & 8\, K\,\lambda\, y\, (2-y)\, \Im m \,C^{\cal I}_{{\rm unp}} \,, \nonumber\\
c_{0,{\rm unp}}^{{\rm BH}} & = & 8\,K^2 \left[(2+3 \epsilon^2) \frac{Q^2}{t} \left(F_{1p}^2-
\frac{t}{4 m_N^2} F_{2p}^2\right)+2 x_B^2 \left(F_{1p}+F_{2p}\right)^2 \right]
\nonumber\\
&+&(2-y)^2 \Bigg\{(2+\epsilon^2)\left[\frac{4 x_B^2 m_N^2}{t} 
\left(1+\frac{t}{Q^2} \right)^2 + 4(1-x_B) \left(1+x_B \frac{t}{Q^2}\right)
\right]\left(F_{1p}^2-
\frac{t}{4 m_N^2} F_{2p}^2\right) \nonumber\\
&+&4x^2\left[x_B+\left(1-x_B+\frac{\epsilon^2}{2}\right)\left(1- \frac{t}{Q^2}\right)^2
-x_B(1-2x_B) \frac{t^2}{Q^4} \right]\left(F_{1p}+F_{2p}\right)^2 \Bigg \} \nonumber\\
&+&8 (1+\epsilon^2) \left(1-y-\frac{\epsilon^2 y^2}{4} \right)
\Big[2 \epsilon^2 \left(1-\frac{t}{4 m_N^2} \right)\left(F_{1p}^2-
\frac{t}{4 m_N^2} F_{2p}^2\right) \nonumber\\
& - & x_B^2 \left( 1-\frac{t}{Q^2}\right)^2 
\left(F_{1p}+F_{2p}\right)^2 \Big]
\,, \nonumber\\
c_{1,{\rm unp}}^{{\rm BH}} & = & 8\,K\,(2-y)\Bigg\{\left(\frac{4 x_B^2 m_N^2}{t}-2x_B-\epsilon^2
 \right) \left(F_{1p}^2-
\frac{t}{4 m_N^2} F_{2p}^2\right) \nonumber\\
&+&2x_B^2 \left(1-(1-2x_B)\frac{t}{Q^2} \right)
\left(F_{1p}+F_{2p}\right)^2 \Bigg\}
\ \,,
\nonumber\\
c_{2,{\rm unp}}^{{\rm BH}} & = & 8\,x_B^2\,K^2 \left\{\frac{4 m_N^2}{t}
\left(F_{1p}^2-
\frac{t}{4 m_N^2} F_{2p}^2\right)+2 \left(F_{1p}+F_{2p}\right)^2 \right\}
 \,,
\nonumber\\
c_{0,{\rm unp}}^{{\rm DVCS}} & = & 2(2-2y+y^2)\,  C^{\rm DVCS}_{{\rm unp}}\,,
\label{eq:a3}
\end{eqnarray}
where 
\begin{eqnarray}
C^{\cal I}_{{\rm unp}} & = & F_{1p} \,{\cal H}_p-\frac{t}{4 m_N^2} F_{2p}\, {\cal E}_p
\,, \nonumber\\
\Delta C^{\cal I}_{{\rm unp}} & = & -\frac{x_B^2}{(2-x_B)^2} 
\left(F_{1p}+F_{2p}\right) \left({\cal H}_p+{\cal E}_p\right) \,, \nonumber\\
C^{\rm DVCS}_{{\rm unp}} & = & \frac{1}{(2-x_B)^2} \Bigg[
4 (1-x_B) |{\cal H}_p|^2-x_B^2 ({\cal H}_p^{\ast} {\cal E}_p+{\cal E}_p^{\ast} {\cal H}_p)
\nonumber\\
&-&\left(x_B^2+(2-x_B)^2\frac{t}{4 m_N^2}\right) |{\cal E}_p|^2 \Bigg]
\label{eq:a4}
\end{eqnarray}
Equations~(\ref{eq:a3}) and (\ref{eq:a4}) involve proton Compton form factors (CFFs)
${\cal H}_p$ and ${\cal E}_p$ and electromagnetic form factors $F_{1p}$ and 
$F_{2p}$. For the CFFs, we used the dual parameterization
with $J_u=J_d=0$~\cite{Guzey:2006xi}.
The proton electromagnetic form factors are parameterized in the following form~\cite{Belitsky:2001ns}
\begin{eqnarray}
F_{1p}(t) & = & \frac{1-(1+k_p)\frac{t}{4 m_N^2}}{1-\frac{t}{4 m_N^2}} G_D(t) \,, \nonumber\\
F_{2p}(t) & = & \frac{k_p}{1-\frac{t}{4 m_N^2}} G_D(t) \,, \nonumber\\
G_D(t) & = & \frac{1}{1-\frac{t}{m_V^2}} \,,
\label{eq:proton_em}
\end{eqnarray}
where $k_p$ is the proton anomalous magnetic moment, $k_p=1.79$; 
$m_V=0.84$ GeV.
More elaborate parameterizations of the nucleon elastic form factors are possible, see e.g.~\cite{Bradford:2006yz}, but Eq.~(\ref{eq:proton_em}) is sufficiently accurate
for our purposes.

\subsection{Neutron part}

Expressions for the $\cos \phi$ and $\sin \phi$-harmonics for the neutron case
are readily obtained from Eqs.~(\ref{eq:a3}) and (\ref{eq:a4})  
by replacing the proton CFFs and electromagnetic form factors by their neutron
counterparts. The neutron CFFs are obtained from the proton ones 
by exchanging $e_u \leftrightarrow e_d$ in the DVCS amplitude.

The 
neutron electromagnetic form factors are parameterized in the following form~\cite{Belitsky:2001ns}
\begin{eqnarray}
F_{1n}(t) & = & -\frac{t}{4 m_N^2}\frac{k_n}{1-\frac{t}{4 m_N^2}} G_D(t) \,, \nonumber\\
F_{2n}(t) & = & \frac{k_n}{1-\frac{t}{4 m_N^2}} G_D(t) \,, 
\end{eqnarray}
where $k_n$ is the neutron anomalous magnetic moment, $k_n=-1.91$.

\end{document}